\documentclass[12pt]{iopart}
\usepackage{iopams}
\expandafter\let\csname equation*\endcsname\relax
\expandafter\let\csname endequation*\endcsname\relax
\usepackage{amsmath,amssymb,bm,yfonts}
\usepackage[pdftex]{hyperref}
\usepackage{cite}
\usepackage[pdftex]{graphicx}
\usepackage[utf8]{inputenc}

\newcommand{\ts}{\textstyle}
\newcommand{\M}{\mathcal M}
\newcommand{\re}{\mathfrak{Re}}
\newcommand{\im}{\mathfrak{Im}}

\makeatletter
\g@addto@macro\@floatboxreset\centering
\makeatother

\begin{document}

\title[Singularities in Reissner-Nordstr\"om black holes]
{Singularities in Reissner-Nordstr\"om black holes}


\author{Paul~M.~Chesler}
\ead{pchesler@g.harvard.edu}

\vspace{10pt}

\address{Black Hole Initiative, Harvard University, Cambridge, MA 02138, USA}

\author{Ramesh Narayan}
\ead{rnarayan@cfa.harvard.edu}

\vspace{10pt}

\address{Center for Astrophysics $|$ Harvard \& Smithsonian, 60 Garden Street, Cambridge, MA 02138, USA and Black Hole Initiative, Harvard University, Cambridge, MA 02138, USA }

\vspace{10pt}

\author{Erik~Curiel}

\address{Munich Center for Mathematical Philosophy, Ludwig-Maximilians-Universit\"at,  
	Ludwigstra\ss \ 31, 80539 M\"unchen, Germany
and Black Hole Initiative, Harvard University, Cambridge, MA 02138, USA}

\vspace{10pt}

\ead{erik@strangebeautiful.com}

\vspace{10pt}

\begin{indented}
\item[] \today
\end{indented}

\begin{abstract}
	We study black holes produced via collapse of a spherically symmetric charged scalar field in asymptotically flat space.  We employ a late time expansion and argue that
	decaying fluxes of radiation through the event horizon imply that the black hole must contain a null singularity on the Cauchy horizon and a central spacelike singularity.
\end{abstract}

%
%
%
%
%

\section{Introduction and summary}

It is widely believed that long after black holes form their exterior geometry is described by the Kerr-Newman metric.  The Kerr-Newman geometry
naturally provides a mechanism for exterior perturbations to relax.  Namely, 
perturbations are either absorbed by the black hole or radiated to infinity. 
Deep inside the black hole though, no such relaxation mechanism exists and the geometry 
depends on initial conditions.

\begin{figure}[h]
	\includegraphics[trim= 0 0 0 0 ,clip,scale=0.4]{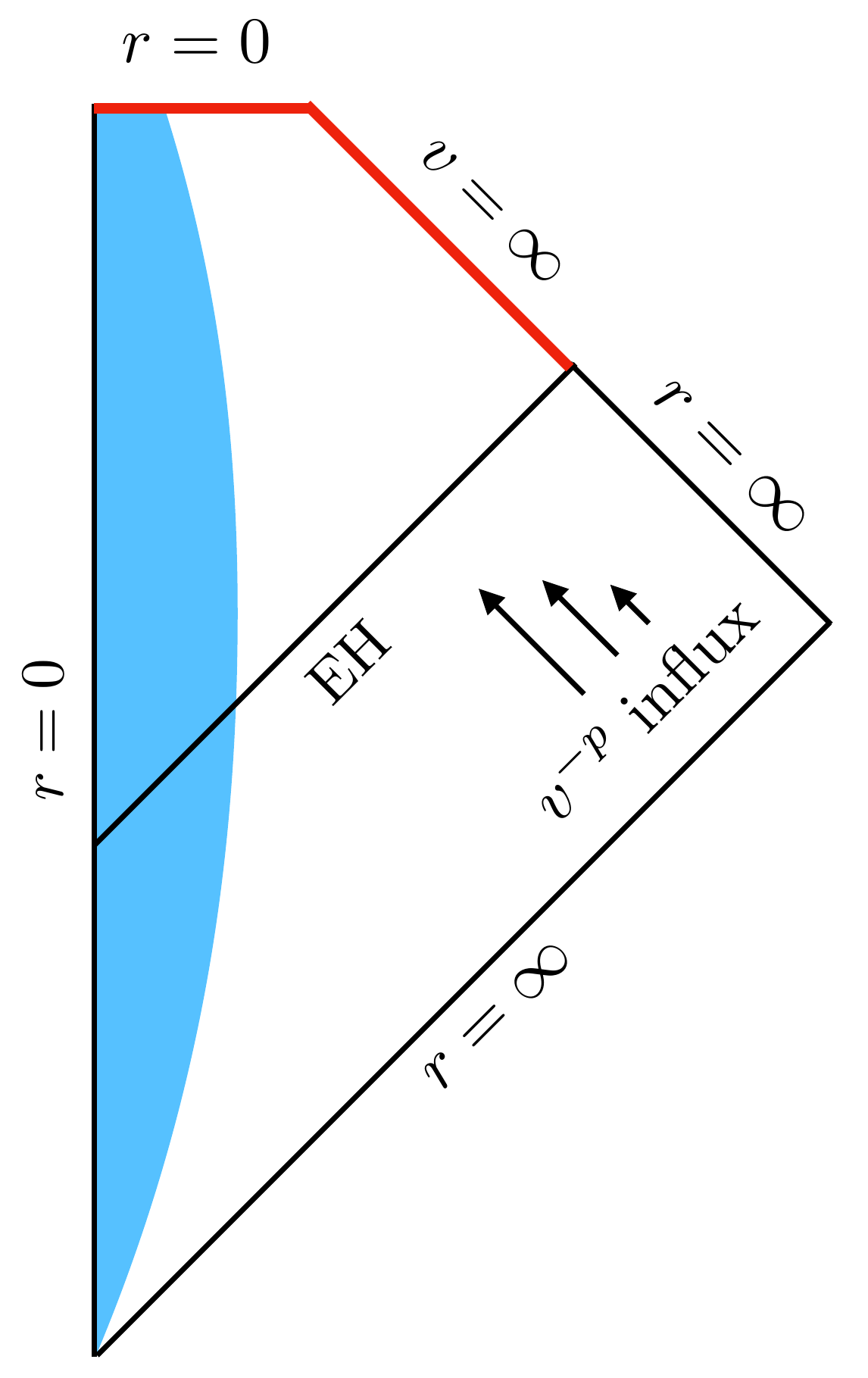}
	\caption{
		A Penrose diagram showing gravitational collapse of matter (blue shaded region) resulting
		in the formation of an event horizon (EH).  There is a decaying 
		flux $\sim v^{-p}$ of infalling radiation through the EH.  The solid red lines  denote singularities.  There is a null curvature singularity on the Cauchy horizon, located at time $v = \infty$, and a spacelike singularity located at $r = 0$.
	}
	\label{fig:PenroseDiagram}
\end{figure}

While the geometry inside the black hole is not unique, it is natural to ask whether there are any 
universal features, such as the structure of singularities.  Consider a black hole produced via
gravitational collapse in asymptotically flat space, such as that shown in the Penrose diagram in Fig.~\ref{fig:PenroseDiagram}.
The collapsing body -- the blue shaded region -- results in an event horizon (EH) forming.  In accord with Price's Law \cite{Price:1971fb,Price:1972pw}, 
collapse also results in an influx of radiation through the EH which decays with an inverse power $v^{-p}$ of advanced time $v$.
Penrose reasoned more than 50 years ago \cite{Penrose:1968ar} that infalling radiation will be infinitely blue shifted 
at the geometry's Cauchy Horizon (CH), located at $v = \infty$, leading to a singularity there.  For Reissner-Nordstr\"om (RN) black holes it was subsequently argued by Poisson and Israel \cite{Poisson:1989zz,PhysRevD.41.1796} that curvature scalars blow up like $e^{2 \kappa v}$, where $\kappa$ is the surface gravity of the inner horizon of the  associated RN solution, leading to a null singularity on the CH.  Numerous studies \cite{Simpson:1973ua,HISCOCK1981110,PhysRevD.20.1260,Poisson:1989zz,PhysRevD.41.1796,PhysRevLett.67.789,0264-9381-10-6-006,Brady:1995ni,Burko:1997zy,Hod:1998gy,Burko:1997fc,10.2307/3597235,Dafermos:2017dbw,Ori:2001pc,Ori1997,Burko:2016uvr,Dias:2018ynt} suggest that the presence of a null singularity at the CH is a generic feature of black hole interiors \footnote{A notable exception are near-extremal black holes in de Sitter spacetime  \cite{Cardoso:2017soq}.}.  

For small spherically symmetric perturbations of two-sided black holes, the null singularity on the CH 
can be the only singularity \cite{Dafermos:2012np}.  However, with large spherically symmetric perturbations, numerical simulations indicate that, in addition to a singular CH, a space-like singularity
forms at areal radius $r = 0$ \cite{Brady:1995ni,Burko:1997zy}. Likewise, for spherically symmetric one-side black holes, which form from gravitational collapse, numerical simulations also indicate 
the formation of a spacelike singularity at $r = 0$ and a singular CH 
\cite{Hod:1998gy}.  This means the singular structure of the spacetime is that shown in Fig.~\ref{fig:PenroseDiagram}.
 
In this paper we focus primarily on one-side black holes (although we will discuss generalizations of our analysis to two-sided black holes in Sec.~\ref{sec:conclusion}). We argue that the formation of a central spacelike singularity is inevitable in the collapse of a spherically symmetric charged scalar field in asymptotically flat spacetime.  
Our analysis employs three key assumptions.  Firstly, we assume Price's Law applies, meaning there is an influx of scalar radiation through the horizon
which decays like $v^{-p}$ for some power $p$ which is sufficiently large such that the mass $M(v)$ and charge $Q(v)$ of the black hole approach constants $M(\infty)$ and $Q(\infty)$ as $v \to \infty$. The infalling radiation (which is left-moving in the Penrose diagram in Fig.~\ref{fig:PenroseDiagram}) 
can scatter off the gravitational field and excite outgoing radiation (which is right-moving in the Penrose diagram).  This, together with outgoing radiation emitted during collapse, means that the interior geometry of the black hole is filled with outgoing radiation.

Our second assumption is that the geometry at areal radius $r >r_-$ relaxes to the RN solution as $v \to \infty$.  The radii $r_\pm$ are the inner $(-)$ and outer $(+)$ horizon radii of the RN solution with mass $M(\infty)$ and charge $Q(\infty)$ \footnote
{
	Recall that the surface $r=r_-$ is null for the RN solution.
	This need not be the case out of equilibrium.  Indeed, prior to collapse the surface $r = r_-$ is time-like.  Our analysis below 
	implies that $r = r_-$ is spacelike at late times.
},
\begin{equation}
\label{eq:rminus}
r_\pm \equiv M(\infty) \pm \sqrt{M(\infty)^2 - Q(\infty)^2}.
\end{equation}
Why is it reasonable to assume the geometry at $r> r_-$ relaxes to the 
RN solution? In the RN geometry, all light rays at $r < r_+$, propagate to $r \leq r_-$ as $v \to \infty$. Therefore, the RN geometry naturally provides a mechanism for perturbations at $r > r_-$ to relax.  
To illustrate this further, in \ref{app} we present a numerically generated solution to the equations of motion, Eqs.~(\ref{eq:einsteinscalar}) below. 

Our third assumption is that at any fixed time $v$, the geometry at $r > 0$ contains no singularities.  This assumption means the equations of motion can be integrated in all the way to $r > 0$ without running into a singularity.
We note, however, that this assumption is {inconsistent} with the weakly perturbed two-sided black holes studied in \cite{Dafermos:2012np}, where at finite time $v$, the outgoing branch of the singular CH lies at $r \approx r_-$.  However, for one-sided black holes, numerical simulations indicate no singularities at $r > 0$ at finite $v$ \cite{Hod:1998gy}. Additionally, our numerical simulations in the Appendix also show no signs of singularities at $r > 0$ at finite $v$.

The assumption that the geometry at $r>r_-$ relaxes to the RN solution
has profound consequences for late-time infalling observers passing through $r_-$. Firstly, as $v \to \infty$ outgoing radiation inside the black hole must be localized to a ball whose surface approaches $r_-$.  This follows from the fact that in the RN geometry, all outgoing light rays between $r_- < r < r_+$ approach $r_-$ as $v \to \infty$.  
Moreover, from the perspective of infalling observers, the outgoing radiation appears blue shifted by a factor of $e^{\kappa v}$.
This means that late-time infalling observers encounter an effective ``shock" at $r = r_-$ \cite{Marolf:2011dj,Eilon:2016osg,Chesler:2018hgn,Burko:2019fgt}, where there is a searing ball of blue shifted radiation.
In particular,  upon passing through $r_-$, infalling observers will measure a Riemann tensor of order $e^{2 \kappa v}$ and therefore experience exponentially large gravitational and tidal forces.
Via Raychaudhuri's equation, the ball of outgoing 
radiation focuses infalling null light rays from $r = r_-$ to $r = 0$ over an affine parameter interval \cite{Marolf:2011dj}
\begin{equation}
\label{eq:Dlambda}
\Delta \lambda \sim e^{-\kappa v}.
\end{equation}
The scaling (\ref{eq:Dlambda}) has been verified numerically for spherically symmetric charged black holes
\cite{Eilon:2016osg} and for rotating black holes \cite{Chesler:2018hgn}.

The exponential focusing of infalling geodesics suggests that at late times there
exists an expansion parameter $\epsilon \equiv e^{-\kappa v} \ll 1$ 
in terms of which the equations of motion can be solved perturbatively in the region $r<r_-$.
This can be made explicit by employing the affine parameter $\lambda$ of infalling null geodesics as a radial coordinate.  With spherical symmetry the metric takes the form
\begin{equation}
\label{eq:metric}
ds^2 = -2 A dv^2 + 2 d\lambda dv + r^2  (d \theta^2 + \sin^2 \theta d\phi^2 ),
\end{equation}
where $\{\theta,\phi\}$ are polar and azimuthal angles respectively. Both $A$ and the areal coordinate $r$ depend on $v$ and $\lambda$.
In this coordinate system curves with $dv= 0$ are radial infalling null geodesics affinely parameterized by $\lambda$.  Shocks at $r = r_-$ then imply derivatives w.r.t. $\lambda$ must diverge like $e^{\kappa v}$ in the region $r < r_-$.  This means that inside $r_-$, the equations of motion can be expanded in powers of $\lambda$ derivatives (\textit{i.e.} a derivative expansion).  This is simply an expansion in powers of $\epsilon$, which is exponentially small as $v \to \infty$. 

\begin{figure}[h]
	\includegraphics[trim= 0 0 0 0 ,clip,scale=0.6]{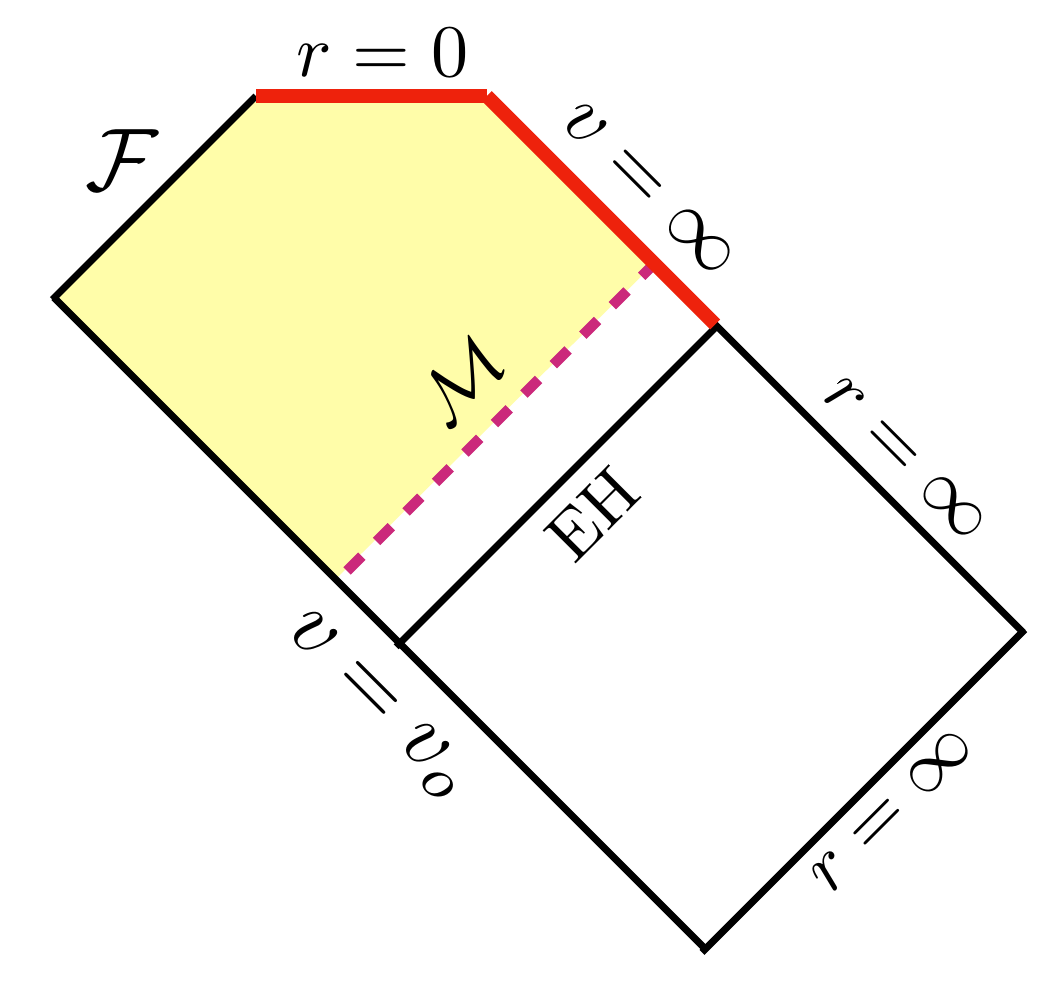}
	\caption{
		A Penrose diagram showing the setup of our problem.  We specify initial data on some infalling $v = v_o$ null surface and boundary data on some outgoing null surface $\M$.
		In the shaded region we solve the equations of motion with a derivative expansion in $\lambda$.  The outgoing null surface $\mathcal F$ bounds the inner domain of validity 
		of the derivative expansion, and always intersects $r = 0$ at some finite time $v$.
		Decaying $v^{-p}$ infalling radiation through $\M$ always results in a null singularity at $v = \infty$
		and a spacelike singularity at $r = 0$, irrespective of initial data at $v = v_o$.
	}
	\label{fig:CompDomain}
\end{figure}

With the metric ansatz (\ref{eq:metric}), initial data is naturally specified on some $v = v_o$ null surface.  We consider the limit where $v_o$ is arbitrarily large.  At $v>v_o$ we restrict our attention to the region inside two outgoing null surfaces $\M$ and $\mathcal F$, as depicted in Fig.~\ref{fig:CompDomain}.  On $\M$ we impose the boundary condition that there is an influx of scalar radiation decaying like $v^{-p}$.  In the shaded region between $\M$ and $\mathcal F$ we solve the equations of motion with a derivative expansion in $\lambda$.  Why have we bothered to introduce $\mathcal F$?  Why not just integrate deeper into the geometry?  It turns out the surface $\mathcal F$ bounds the inner domain of validity of the derivative expansion: in the region enclosed by $\mathcal F$ the derivative expansion can break down.  However, as depicted in Fig.~\ref{fig:CompDomain}, we find that $\mathcal F$ propagates \textit{inwards} and intersects $r = 0$ at a finite time $v$.  
In other words, the domain of dependence and validity of the derivative expansion initial value problem contains $r = 0$ at late enough times.

We find that infalling radiation through $\M$ results in the Kretchmann scalar diverging like $e^{2 \kappa v}$, consistent with previous demonstrations of a singular CH \cite{Simpson:1973ua,HISCOCK1981110,PhysRevD.20.1260,Poisson:1989zz,PhysRevD.41.1796,PhysRevLett.67.789,0264-9381-10-6-006,Brady:1995ni,Burko:1997zy,Hod:1998gy,Burko:1997fc,10.2307/3597235,Dafermos:2017dbw,Ori:2001pc,Ori1997,Burko:2016uvr}.  Additionally, we find that infalling radiation results in a cloud of radiation forming near $r = 0$.  This cloud always results in a spacelike singularity 
forming at $r = 0$ at late enough times, irrespective of initial conditions at $v = v_o$.
In particular, the growing cloud of radiation results in the Kretchmann scalar diverging near $r = 0$ like $r^{-2 \alpha v} e^{2\kappa v}$ for some constant $\alpha > 0$.

An outline of the rest of the paper is as follows.  In Sec.~\ref{sec:eqm} we
write the equations of motion, employing the affine parameter $\lambda$ as a radial coordinate.  In Sec.~\ref{sec:derexpansion} we present the leading order equations of motion within the derivative expansion. In Sec.~\ref{sec:singularstructRN} we employ the leading order equations of motion to study the causal structure of singularities inside the black hole, and in Sec.~\ref{sec:conclusion} we present concluding remarks.
In \ref{app}, as an example, we present a numerical solution to the equation of motion.

\section{Equations of motion}
\label{sec:eqm}

We consider the dynamics of 
spherically symmetric charged black holes with a charged scalar field $\Psi$ and a gauge field $\mathcal A_\mu$.
The equations of motion read
\begin{align}
	\label{eq:eqm}
	&R_{\mu \nu} + {\textstyle \frac{1}{2}} R g_{\mu \nu} = 8 \pi (T_{\mu \nu} + \tau_{\mu \nu}), &
	&D_\mu F^{\mu \nu} = J^\nu, &
	&\mathcal D^2 \Psi = 0,&
\end{align}
where $D_\mu$ is the covariant derivative, $\mathcal D_\mu = D_\mu - i \mathcal A_\mu$ is the gauge covariant derivative, and
\begin{subequations}
	\label{eq:stresses}
\begin{align}
	T_{\mu \nu} &= 2 \re \{ (\mathcal D_\mu \Psi)^* \mathcal D_\nu \Psi\} - g_{\mu \nu} |\mathcal D \Psi|^2, 
	\\
	\tau_{\mu \nu} &=-F_{\mu \beta}F^{\beta}_{\ \nu} - {\textstyle \frac{1}{4}} g_{\mu \nu} F_{\alpha \beta}F^{\alpha \beta},
	\\
	 J^\mu &= -2 \im \{\Psi^*\mathcal D^\mu \Psi\},
\end{align}
\end{subequations}
are the scalar and electromagnetic stress 
tensors and electric current, respectively.

We work in the gauge $\mathcal A_{\lambda} = 0$.  With the metric ansatz (\ref{eq:metric}), the components of the equation of motion (\ref{eq:eqm}) then read
\begin{subequations}
	\label{eq:einsteinscalar}
	\begin{align}
	\label{eq:reqRN}
	0 &=r'' + 8 \pi |\Psi'|^2 r,
	\\ \label{eq:rdoteqRN}
	0 &= (r \, d_+ r)' {\textstyle +2 \pi r^2 E^2 - \frac{1}{2}},
	\\ \label{eq:AeqRN}
	0&= A''  {\ts - \frac{2 r'}{r^2} d_+ r + \frac{1}{r^2}+ 8 \pi \left ( 2 \, \re \{ \Psi^{'*} \mathcal D_+ \Psi \} -  E^2 \right) },
	\\ \label{eq:rddotRN}
	0&= d_+^2 r  - A' d_+ r + 8 \pi r \, |\mathcal D_+ \Psi|^2,
	\\ \label{eq:scalareqRN}
	0&= \ts (r \, \mathcal D_+ \Psi)' + \Psi' d_+ r - \frac{i}{2} E \Psi,
	\\ \label{eq:EeqRN}
	0&= (r^2 E)'- 2 r^2 \im \{ \Psi^* \Psi'\}, 
	\\ \label{eq:A0constraint}
	0&= \ts d_+ E + \frac{2 E}{r}  d_+ r - 2\im \{ \Psi^* \mathcal D_+ \Psi\},
	\end{align}
\end{subequations}
where 
\begin{equation}
\label{eq:electricfield}
E \equiv -\mathcal A_0',
\end{equation}
is the electric field.  The derivative operators $'$, $d_+$ and $\mathcal D_+$ are defined to be
\begin{align}
&' \equiv \partial_\lambda,&& d_+ \equiv \partial_v + A \partial_\lambda,&& \mathcal D_+ = d_+ - i \mathcal A_0.
\end{align} 
The $'$ derivative is just the directional derivative along
ingoing null geodesics whereas $d_+$ is the directional derivative 
along outgoing null geodesics.  $\mathcal D_+$ is simply the gauge covariant
version of $d_+$.

Eq.~(\ref{eq:reqRN}) is an initial value constraint: if (\ref{eq:reqRN}) is satisfied at $v = v_o$, then the remaining equations guarantee it will remain satisfied 
at later times.  Eqs.~(\ref{eq:rddotRN}) and (\ref{eq:A0constraint}) are
radial constraint equations: if (\ref{eq:rddotRN}) and (\ref{eq:A0constraint}) are satisfied at one value of $\lambda$, the remaining equations guarantee they will remain satisfied at all values of $\lambda$.  

The equations of motion (\ref{eq:einsteinscalar}) constitute a nested system of linear ODEs.  Given $\Psi$ at $v = v_o$ and boundary data on the outgoing null surface $\M$, shown in the Penrose diagram in Fig.~\ref{fig:CompDomain}, Eq.~(\ref{eq:reqRN}) can be integrated in from $\M$ to find $r$.  Next, given $\Psi$ and $r$, Eq.~(\ref{eq:EeqRN}) 
can be integrated in from $\M$ to find $\mathcal A_0$.
With $\Psi$, $r$ and $\mathcal A_0$ known, Eq.~(\ref{eq:rdoteqRN}) can be integrated in from $\M$ to find $d_+ r$.  With $\Psi$, $r$, $\mathcal A_0$ and $d_+ r$ known, Eq.~(\ref{eq:scalareqRN}) can be integrated in to find $ \mathcal D_+ \Psi$.  With $\Psi$, $r$, $\mathcal A_0$, $d_+ r$ and $\mathcal D_+ \Psi$ known, Eq.~(\ref{eq:AeqRN}) can be integrated in from $\M$ to find $A$.  
With $\Psi$, $\mathcal A_0$, $A$ and $\mathcal D_+ \Psi$ known, we can compute $\partial_v \Psi$ and advance forward in time.
Note the remaining equations, Eqs.~(\ref{eq:rddotRN}) and (\ref{eq:A0constraint}), which are radial constraint equations, can be implemented as boundary conditions in the aforementioned radial integrations.

\section{Derivative expansion}
\label{sec:derexpansion}

Following our arguments in the Introduction, in the region enclosed by the outgoing null surface $\M$ we shall solve the equations of motion (\ref{eq:einsteinscalar})
with a derivative expansion in $\lambda$.  
For pedagogical reasons we choose $\M$ to asymptote to $r = r_-$ as $v \to \infty$.  However, it will turn out that the 
precise choice of $\M$ doesn't matter for our analysis.
One could equally well choose $\M$ to asymptote to some finite $r < r_-$.

The metric (\ref{eq:metric}) is invariant under the residual diffeomorphism 
\begin{equation}
\label{eq:resdiff}
\lambda \to \lambda  + \xi(v),
\end{equation}
where $\xi(v)$ is arbitrary.  
We exploit this residual diffeomorphism invariance to choose coordinates such that $\M$ lies at $\lambda = 0$, with the spacetime enclosed by $\M$ lying at $\lambda < 0$.  Since $\M$ is null, this means
\begin{equation}
\label{eq:AvanishingRN}
A|_{\lambda = 0} = 0.
\end{equation}
Additionally, the gauge choice $\mathcal A_\lambda = 0$ enjoys the residual gauge freedom
\begin{equation}
\label{eq:resgauge}
\mathcal A_0 \to \mathcal A_0 +  \Lambda(v), 
\end{equation}
where $\Lambda(v)$ is arbitrary. We exploit this residual gauge freedom 
to set 
\begin{equation}
\label{eq:A0vanishingRN}
\mathcal A_0|_{\lambda = 0} = 0.
\end{equation}

In order to account for rapid $\lambda$ dependence, 
we introduce a bookkeeping parameter $\epsilon$ and assume the following 
scaling relations for the directional derivatives along infalling and outgoing null geodesics,
\begin{align}
\label{eq:scalings}
&\partial_\lambda = O(1/\epsilon), && d_+ = O(\epsilon^0).
\end{align}
We then study the equations of motion (\ref{eq:einsteinscalar}) in the $\epsilon \to 0$ limit.  We shall see below that as advertised in the Introduction, $\epsilon = e^{-\kappa v}$.  Hence the $\epsilon \to 0$ is 
just the $v\to \infty$ limit.

Why must we have $d_+ = O(\epsilon^0)?$
For any quantity $f$, the total $v$ derivative of $f$ along outgoing geodesics is $\frac{df}{dv} = d_+ f$.
Hence the scaling $d_+ = O(\epsilon^0)$ reflects the fact that 
quantities evaluated along outgoing null geodesics are not rapidly varying in $v$.
The scalings (\ref{eq:scalings}) and Gauss' law (\ref{eq:EeqRN}) also imply $E = O(\epsilon^0)$. Eq.~(\ref{eq:electricfield}) and the boundary condition (\ref{eq:A0vanishingRN}) then imply
\begin{align}
\mathcal A_0 = O(\epsilon).
\end{align}
Likewise, the scaling relations (\ref{eq:scalings}) and the Einstein equation (\ref{eq:AeqRN}) imply $A'' = O(1/\epsilon)$.  Together with the boundary condition (\ref{eq:AvanishingRN}), this means
\begin{align}
A = O(\epsilon).
\end{align}

Further boundary conditions are needed on $\M$.  Firstly, we assume the influx of scalar radiation through $\M$ is a power law in accord with Price's Law:
\begin{equation}
\label{eq:influxRN}
\mathcal D_+ \Psi|_{\lambda = 0} =  d_+ \Psi|_{\lambda = 0} \sim v^{-p}.
\end{equation}
We leave 
$p$ arbitrary. Second, we fix a Neumann boundary condition on $A$.  
The scaling $A' = O(\epsilon^0)$ implies $\partial_r (A') = O(\epsilon^0)$.  Hence it is reasonable to assume $A'$ remains continuous in $r$ across $r_-$ as $\epsilon \to 0$, or equivalently as $v \to \infty$.   We note this assumption is consistent with the numerical simulation presented in  \ref{app}. Additionally, we note this assumption is also constant with numerical solutions of the interior of rotating black holes \cite{Chesler:2018hgn}. 
With our assumption that the geometry at $r > r_-$ relaxes to the RN solution as $v \to \infty$, this means $A'|_{\lambda = 0}$ must approach its RN limit,
\begin{equation}
\label{eq:surfacegravityRN}
A'|_{\lambda = 0} = -\kappa,
\end{equation}
with $\kappa$ the surface gravity of the associated Reissner-Nordstr\"om inner horizon,
\begin{equation}
\kappa = \frac{Q(\infty)^2 - M(\infty) r_-}{r_-^3}.
\end{equation}

In the $\epsilon \to 0$ limit, Eqs.~(\ref{eq:rdoteqRN})--(\ref{eq:scalareqRN}) and (\ref{eq:A0constraint}) read
\begin{subequations}
	\label{eq:einsteinscalar2}
	\begin{align}
	\label{eq:rdoteqRN2}
	0 &= (r \, d_+ r)',
	\\ \label{eq:AeqRN2}
	0&= A''  {\ts - \frac{2 r'}{r^2} d_+ r + 16 \pi \, \re \{ \Psi^{'*} d_+ \Psi \} },
	\\ \label{eq:rddotRN2}
	0&= d_+^2 r  - A' d_+ r + 8 \pi r \, | d_+ \Psi|^2,
	\\ \label{eq:scalareqRN2}
	0&= (r \,  d_+ \Psi)' + \Psi' d_+ r,
	\\ \label{eq:A0constraint2}
	0&= \ts d_+ E + \frac{2 E}{r}  d_+ r - 2\im \{ \Psi^*  d_+ \Psi\}.
	\end{align}
\end{subequations}
The remaining equations of motion (\ref{eq:reqRN}) and (\ref{eq:EeqRN}) do not change in the $\epsilon \to 0$ 
limit.  

Eq.~(\ref{eq:rdoteqRN2}) can be integrated to yield
\begin{equation}
\label{eq:rdotRN}
d_+ r = -\frac{\zeta(v)}{r}.
\end{equation}
The constant of integration $\zeta(v)$ can be determined from the radial constraint equation (\ref{eq:rddotRN2}).  Consider this equation
evaluated on $\M$.
Since $d_+$ is the 
directional derivative along outgoing null geodesics,  Eq.~(\ref{eq:rddotRN2}) can be rewritten on $\M$ as an ODE for $d_+ r$,
\begin{equation}
\label{eq:constraintlove}
\frac{d}{dv} d_+ r + \kappa d_+ r =  - 8 \pi r | d_+ \Psi|^2,
\end{equation}
where we have employed (\ref{eq:surfacegravityRN}) to eliminate $A'$.
Employing Price's law (\ref{eq:influxRN}), in the $v \to \infty$ limit
this equation is solved by (\ref{eq:rdotRN}) with
\begin{equation}
\label{eq:zetaRN}
\zeta(v) \sim +v^{-2p}.
\end{equation}

We emphasize that $d_+r < 0$.  This means there is no inner apparent horizon inside $\M$, for 
at an apparent horizon $d_+r =0$.
Since $d_+ r$ is the directional derivative of $r$ along outgoing null geodesics, Eqs.~(\ref{eq:rdotRN}) and (\ref{eq:zetaRN}) yield the outgoing geodesic equation  
\begin{equation}
\label{eq:outgoinggeos}
\frac{dr}{dv} = d_+ r \sim - \frac{v^{-2p}}{r}.
\end{equation}
Eq.~(\ref{eq:outgoinggeos}) can be integrated to yield
\begin{equation}
\label{eq:outgeosol}
r^2 \sim v^{1 - 2 p} + {\rm const.}
\end{equation}
It follows that geodesics with $r \leq r_{\rm c}$
end at $r = 0$ whereas those with $r > r_{\rm c }$ end on the CH at finite values of $r$.  This is consistent 
with the Penrose diagrams in Figs.~\ref{fig:PenroseDiagram} and \ref{fig:CompDomain}.
The critical radius $r_{\rm c}$ is given by 
\begin{equation}
\label{eq:logkicksin}
r_{\rm c}(v)  \sim v^{1/2 - p}.
\end{equation}

The remaining equations of motion (\ref{eq:AeqRN2}) and (\ref{eq:scalareqRN2}) cannot be solved analytically without further approximations.  In Sec.~\ref{sec:nearMRN} we shall solve these equations near $\M$, where infalling radiation can be treated perturbatively, and establish the self-consistency condition that $\lambda$ derivatives indeed blow up like $e^{\kappa v}$.  In Sec.~\ref{sec:nearoriginRN}
we shall show that near $r = 0$, $\lambda$ derivatives
blow up like $r^{ -\alpha v} e^{\kappa v}$ where $\alpha > 0$ is some constant. 
In Sec.~\ref{sec:consistancyRN} we discuss the domain of validity of the derivative expansion.

\subsection{Derivative expansion near $\M$}
\label{sec:nearMRN}

In this section we solve Eqs.~(\ref{eq:einsteinscalar})
near $\M$, meaning away from $r = 0$. 
Since infalling radiation decays as $v \to \infty$, we can neglect its effects near $\M$ at late times.  This is tantamount to imposing the boundary conditions 
$d_+ r = d_+ \Psi = 0$ on $\M$.
In this case Eqs.~(\ref{eq:rdoteqRN2}), (\ref{eq:AeqRN2}) and (\ref{eq:scalareqRN2}) reduce to
\begin{align}
\label{eq:optics}
&d_+ r = 0, &
&d_+ \Psi = 0,&&
 A'' = 0. 
\end{align}
The first two equations here simply state that excitations in $r$ and $\Psi$ are transported
along outgoing null geodesics.
Using the boundary conditions (\ref{eq:AvanishingRN}) and (\ref{eq:surfacegravityRN}), the solutions to (\ref{eq:optics}) read
\begin{align}
\label{eq:outgoingsols}
	&\Psi = \chi(e^{\kappa v}\lambda), &
	&r = \rho(e^{\kappa v}\lambda), &
	&A  = -\kappa \lambda, 
\end{align}
where $\chi$ and $\rho$ are arbitrary 
\footnote{We note, however, that $\chi$ and $\rho$ are related to each other by the initial value constraint (\ref{eq:reqRN}).}. 
The function $\chi$ encodes an outgoing flux
of scalar radiation inside the black hole.  This outgoing radiation need not fall into $r = 0$, just as the Penrose diagram in Fig.~\ref{fig:PenroseDiagram}, suggests. Moreover, 
we see from (\ref{eq:outgoingsols}) that $A' = -\kappa$, even away from $\M$.  We note that this behavior is also see in the numerically generated solution presented in \ref{app}.
It follows that the boundary conditions we imposed on $\M$ are in fact valid in the interior of $\M$, meaning our results are insensitive to the precise choice of $\M$: we could have equally well chosen $\M$
to asymptote to some finite $r < r_-$ as $v \to \infty$.

From (\ref{eq:outgoingsols}) we see that $\lambda$ derivatives blow up like $e^{\kappa v}$. Hence the derivative expansion is simply a late time expansion with expansion parameter $\epsilon \equiv e^{-\kappa v}$. 
Additionally, from (\ref{eq:reqRN}) we see that $r'$ can only \textit{increase} as $\lambda$, or equivalently $r$, decreases.  This means that $\lambda$ derivatives must be \textit{at least} as large as $e^{\kappa v}$ throughout the entire interior of $\M$.

As mentioned above, the physical origin of large $\lambda$ derivatives lies in the fact that from the perspective of infalling observers, outgoing radiation is blue shifted by a factor of $e^{\kappa v}$. 
Moreover, the 
exponentially diverging $\lambda$ derivatives also imply that the Riemann tensor 
diverges like $e^{2 \kappa v}$.  It follows that 
infalling observers experience exponentially large gravitational and tidal 
forces at $r_-$.
This is the gravitational shock phenomenon explored in \cite{Marolf:2011dj,Eilon:2016osg,Chesler:2018hgn}.

\subsection{Derivative expansion near $r = 0$}
\label{sec:nearoriginRN}

The analysis in the preceding section neglected the effects of 
infalling radiation.  However, as can be seen from (\ref{eq:rdotRN}), the amplitude of infalling 
radiation becomes non-negligible as $r \to 0$.  Consequently, its effects must be taken into account 
at small enough $r$.  We show in this section that when infalling radiation is not neglected, at small $r$ derivatives w.r.t. $\lambda$  blow up like $r^{-\alpha v} e^{\kappa v}$ where $\alpha$ is a positive constant.  
How does the $r^{-\alpha v}$ enhancement arise?  A clue comes from 
the initial value constraint (\ref{eq:reqRN}). As mentioned above, from this equation we see that $r'$ can only increase as $\lambda$, or equivalently $r$, decreases.  If the scalar field $\Psi$ diverges 
near $r = 0$, then (\ref{eq:reqRN}) means that $r'$ can diverge there too.   We shall see that such a divergence in $\Psi$ is inevitable at late times due to the influx of scalar radiation through $\M$.

To study the behavior of the scalar field near $r = 0$ we have found
it convenient to change radial coordinates from $\lambda$ to $r$.
In the $(v,r)$ coordinate system we have 
\begin{align}
\label{eq:dplusrsystem}
d_+ &= \partial_v + (d_+ r) \partial_r,
\\ \nonumber
&= \ts \partial_v - \frac{\zeta}{r} \partial_r,
\end{align}
where in the last line we used (\ref{eq:rdotRN}).
The scalar equation of motion (\ref{eq:scalareqRN2})
then reads
\begin{equation}
\label{eq:scalarwaverRN}
\partial_r (r d_+ \Psi) = \frac{\zeta}{r} \partial_r \Psi.
\end{equation}
where we have again used (\ref{eq:rdotRN}) to eliminate $d_+ r$
from the r.h.s. of (\ref{eq:scalareqRN2}).
We therefore reach the conclusion that in the $\epsilon \to 0$ limit the scalar field satisfies a decoupled linear wave equation.

The equation of motion (\ref{eq:scalarwaverRN}) implies the ``energy density"
\begin{align}
\mathcal E \equiv r |\partial_r \Psi|^2, 
\end{align}
satisfies the conservation law
\begin{align}
\label{eq:scalarcons}
\partial_v \mathcal E + \partial_r \mathcal S = 0,
\end{align}
where the flux $\mathcal S$ is given by
\begin{align}
\label{eq:scalarflux}
\mathcal S &\equiv \ts \frac{r^2 }{\zeta}  |d_+ \Psi|^2 - \zeta \, |\partial_r \Psi|^2.
\end{align}
Via Price's law (\ref{eq:influxRN}) and Eq.~(\ref{eq:zetaRN}), the flux through $\M$ scales with $v$ like
\begin{align}
\label{eq:fluxthroughM}
\mathcal S|_{\M} \sim \frac{v^{-2 p}}{\zeta} \sim  1.
\end{align}
Hence, the energy enclosed by $\M$ must increase linearly in $v$.  Moreover, owing to the fact that the explicit time dependence in the equation of motion (\ref{eq:scalarwaverRN}) --- that from $\zeta(v)$ --- is arbitrarily slowly varying at late times, the energy flux must be approximately constant in time throughout the region enclosed by $\M$.

Near the origin Eq.~(\ref{eq:scalarwaverRN}) can be solved 
with a Frobenius expansion,
\begin{equation}
\label{eq:Frobenius}
\Psi(v,r) =  \log r \sum_{n = 0} \Psi_{(n)}(v) {\ts \left (\frac{r}{r_{\rm c}} \right)^{2n}} + \sum_{n = 0} \psi_{(n)}(v) {\ts \left (\frac{r}{r_{\rm c}} \right)^{2n},}
\end{equation}
where $r_c$ is defined in (\ref{eq:logkicksin}).  The condition $\partial_v \mathcal S = 0$ implies 
\begin{equation}
\partial_v \left [ \Psi_{(0)} \partial_v \Psi_{(0)} \right] = 0.
\end{equation}
It follows that as $v \to \infty$ we must have $\Psi_{(0)}(v) \sim \sqrt{v}$.
Evidently, driving the scalar field with a tiny decaying flux $v^{-p}$ of infalling radiation results in the growth of a cloud of scalar radiation at $r \lesssim r_{\rm c}$ with ever increasing radial derivatives as time progresses.  It follows that for 
$r \lesssim r_c$ the energy density $\mathcal E$ must diverge like 
\begin{align}
\label{eq:energyscalings}
\mathcal E \sim \frac{v}{r}.
\end{align} 

Let us now return to using the affine parameter $\lambda$ 
as a radial coordinate and investigate the consequences of 
Eq.~(\ref{eq:energyscalings}) on the behavior of $r'$ as $r \to 0$.  
Eq.~(\ref{eq:reqRN}) can be written
\begin{equation}
\ts r'' + 8 \pi \mathcal E r'^2 = 0.
\end{equation}
Using (\ref{eq:energyscalings}), this equation can be integrated near $r = 0$ to yield $r' \sim r^{-\alpha v} C$ 
where $C$ is a constant of integration and $\alpha > 0$ is a constant.
Recall that near $\M$ we have $r' \sim e^{\kappa v}$ and that 
$r'$ can only increase as $r$ decreases.
It follows that $C \sim e^{\kappa v}$.  We therefore conclude that for $r \lesssim r_c$ we have
\begin{equation}
\label{eq:rscalingenhanced}
r' \sim r^{-\alpha v} e^{\kappa v}.
\end{equation}
Likewise, the chain rule implies that for $r \lesssim r_{\rm c}$ we also have 
\begin{equation}
\label{eq:Psiscaling}
\Psi' \sim r^{-\alpha v} e^{\kappa v}.
\end{equation}

\subsection{Domain of validity of the approximate equations of motion.}
\label{sec:consistancyRN}
 
In deriving the approximate equations of motion (\ref{eq:einsteinscalar2}) we have neglected some terms which can diverge like 
$1/r^q$ near $r = 0$ for some fixed power $q$.  
The neglected divergent terms are a  $1/r^2$ term in Eq.~(\ref{eq:AeqRN}) and terms with the electric field $E$ (which can diverge like $1/r^2$) in Eqs.~(\ref{eq:rdoteqRN}), (\ref{eq:AeqRN}) and (\ref{eq:scalareqRN}).  Suppose initial data is specified at $v = v_o$, as depicted in Fig.~\ref{fig:CompDomain}, with $\lambda$ derivatives initially of order $e^{\kappa v_o}$. If the initial scalar field data is non-singular at $v = v_o$, meaning $\lambda$ derivatives are initially finite at $r = 0$, the approximate 
equations of motion need not be valid beyond the point where the neglected $1/r^q$ terms become comparable to $\lambda$ derivatives.
In other words, the approximate equations of motion can break down when $\log \frac{1}{r} \sim v$. 

However, even with regular initial data, the analysis in the preceding section demonstrates infalling radiation enhances $\lambda$ derivatives at late times by a factor of $r^{-\alpha v}$ for some constant $\alpha > 0$.  The enhancement sets in at $r \sim r_{\rm c}$ and means that $\lambda$ derivatives dominate over any $1/r^q$ divergence at late enough times.  In other words, at late enough times the approximate equations of motion (\ref{eq:einsteinscalar2}) are valid all the way to $r = 0$.

What then is the domain of dependence and validity of the derivative expansion initial value problem?
We can easily bound the inner domain of validity of the approximate 
equations of motion with some outgoing null surface $\mathcal F$, as depicted in Fig.~\ref{fig:CompDomain}.  Define $\mathcal F$ to be the outgoing null surface with initial condition
\begin{equation}
r(v_o)^2 = r_{\rm c}(v_o)^2 (1 - \delta),
\end{equation}
with $0<\delta < 1$. The evolution of $\mathcal F$ is 
governed by the geodesic solution (\ref{eq:outgeosol}) and reads
\begin{equation}
\label{eq:Fdef}
r^2 \sim v^{1 - 2p} - v_o^{1 - 2p} \delta.
\end{equation}
On $\mathcal F$ we can compare the $1/r^q$ divergences to $\lambda$ derivatives. 
Consider first the limit $\delta \to 0$.  In this case $\mathcal F$ coincides with the critical radius $r_{\rm c}$ and intersects $r = 0$ at $v = \infty$.  On $\mathcal F$ the $1/r^q \sim v^{q (2p-1)}$ terms are always parametrically small compared to $e^{\kappa v}$.  In other words, on $\mathcal F$ the $1/r^q$ divergences are always negligible compared to $\lambda$ derivatives, irrespective of the $r^{-\alpha v}$ late-time enhancement.  
Consider then the case where $\delta$ is arbitrarily small but finite.
In this case Eq.~(\ref{eq:Fdef}) implies that $\mathcal F$ intersects 
$r = 0$ at $v = v_*$ where $v_{*}= v_o/ \delta^{1/(2 p - 1)}$.  This means the $1/r^q$ terms diverge on $\mathcal F$ at finite time $v = v_* \gg v_o$. How do the $1/r^q$ divergences compare to $\lambda$ derivatives?  In particular, do $\lambda$ derivatives diverge faster?  The answer is clearly yes due to the $r^{-\alpha v}$ late-time enhancement: the exponent $\alpha v_*$ can be made arbitrarily large by taking $\delta$ smaller whereas the exponent $q$ is fixed.  This means that, as depicted in Fig.~\ref{fig:CompDomain}, the domain of dependence and validity of the derivative expansion initial value problem always contains $r = 0$ at late enough times.

\section{Singular structure inside $\M$}
\label{sec:singularstructRN}

We now explore the consequences of the derivative expansion 
on the structure of singularities inside $\M$.  First we will 
study the singularity on the Cauchy horizon at $v = \infty$.  
Using the equations of motion (\ref{eq:einsteinscalar}) to eliminate second
order derivatives, without approximation the Kretschmann  scalar $K \equiv R^{\mu \nu \alpha \beta} R_{\mu \nu \alpha \beta}$ reads
\begin{align}
\nonumber
K & = \ts 512 \pi^2 \re \{ (\Psi'^* \mathcal D_+ \Psi)^2 \} 
+ 1536 \pi^2 |\Psi' \mathcal D_+ \Psi|^2
+ \frac{48 r'^2 (d_+ r)^2}{r^4} - \frac{256 \pi r' d_+ r}{r^2} \re \{ \Psi'^* \mathcal D_+ \Psi \} 
\\ \label{eq:KRN}
& \ts + \frac{128 \pi(1 - 8 \pi r^2 E^2)}{r^2} \re \{ \Psi'^* \mathcal D_+ \Psi\} - \frac{48( 1 - 4 \pi r^2 E^2)}{r^4} r' d_+ r
+ 320 \pi^2 E^4 + \frac{12}{r^4} - \frac{96 \pi E^2}{r^4}.
\end{align}
From the scaling relations (\ref{eq:rscalingenhanced}) and (\ref{eq:Psiscaling}) we see that the dominant terms 
in $K$ are the four in the first line.  These all blow up like $e^{2 \kappa v}$ as $v \to \infty$.

Let us first focus on the region near $\M$, where we can employ the
solutions (\ref{eq:optics}) for outgoing radiation. Without 
infalling radiation $K$ is regular near $\M$.  However, due to the 
fact that $\lambda$ derivatives blow up exponentially like $e^{\kappa v}$, a tiny amount of 
infalling radiation leads to $K$ growing exponentially.
Accounting for infalling radiation by employing (\ref{eq:rdotRN}), (\ref{eq:zetaRN}) and Price's law (\ref{eq:influxRN}) to determine $d_+ r$ and $\mathcal D_+ \Psi$, we see that the most singular terms in $K$  are the first two, which scale like
\begin{equation}
\label{eq:KdivergRN}
K \sim e^{2 \kappa v} v^{-2 p}.
\end{equation}

The physical origin of the exponential growth (\ref{eq:KdivergRN}) is easy to understand.  Consider 
for the sake of example two scalar wavepackets of wavelength $L_+$ and $L_-$ and amplitude $a_+$ and $a_-$.  When the wavepackets pass through each other the Kretchmann scalar will scale like $(a_+/L_+)^2 (a_-/L_-)^2$.  In the limit 
where either $L_\pm  \to 0$, the \textit{crossing fluxes} of radiation will result in the Kretchmann scalar diverging.  Inside the black hole the solution (\ref{eq:optics}) provides an outgoing flux of scalar radiation while the influx is provided via Price's law, (\ref{eq:influxRN}). The final ingredient is that from the perspective of infalling observers, the outgoing radiation appears blue shifted by $e^{\kappa v}$ (which manifests itself in our coordinate system as $\lambda$ derivatives growing like $e^{\kappa v}$) 
\footnote{From the perspective of \textit{outgoing} observers, ingoing radiation appears blue shifted by $e^{\kappa v}$.}.  This means that the crossing fluxes must result in a scalar curvature singularity on the CH which 
diverges like (\ref{eq:KdivergRN}).

It is interesting to compare (\ref{eq:KdivergRN}) to the contribution 
to the curvature from mass inflation \cite{Poisson:1989zz,PhysRevD.41.1796}.  The mass function 
$m \sim r' d_+r$
and hence, via the second term in (\ref{eq:KRN}),  contributes to $K$ a term $\sim m^2$.  From (\ref{eq:rdotRN}), (\ref{eq:zetaRN}) and (\ref{eq:optics}) we 
see that $m^2 \sim e^{2 \kappa v} v^{-4 p}$.  Hence the contribution
to the curvature from mass inflation is suppressed relative to that of the  crossing fluxes by a factor of $v^{-2 p}$.  Similar results have been 
reported for rotating black holes in \cite{Ori:2001pc}.

We now consider $K$ in the limit $r \to 0$.  Using the scaling relations 
(\ref{eq:rscalingenhanced}) and (\ref{eq:Psiscaling}) we see that the 
$e^{2 \kappa v}$ scaling in (\ref{eq:KdivergRN}) must be enhanced to 
\begin{equation}
\label{eq:Kdivorigin}
K \sim r^{-2 \alpha v} e^{2 \kappa v}
\end{equation}
as $r \to 0$.  In other words, the exponential divergence near the Cauchy horizon becomes stronger when $r \to  0$.  

We now turn to the nature of the singularity at $r = 0$.  From (\ref{eq:Kdivorigin})
we see that the divergence in $K$ near $r = 0$ becomes stronger as $v$ increases.
This is due to the buildup of scalar radiation near $r = 0$ from infalling radiation.  
Moreover, it follows from the outgoing null geodesic solution (\ref{eq:outgeosol}) that 
geodesics at $r < r_{\rm c}$ with $r_{\rm c} \sim v^{1/2 - p}$ must terminate at $r = 0$
in a finite time $v$. This means that the singularity at $r = 0$ must be spacelike.

\section{Concluding remarks}

\label{sec:conclusion}

In this paper we have introduced a novel approximation scheme valid in the interior of black holes which is simply a late-time expansion.  Together with  a set of assumptions, we have employed this scheme to show that decaying $1/v^p$ fluxes of radiation 
through the horizon necessitate the existence of a spacelike singularity at $r =0$.  While we have focused on spherical symmetry, our analysis readily generalizes to geometries with no symmetry including that of rotating black holes.  We shall report on this in a coming paper.

It is interesting that the structure of the singularity at $r = 0$ is independent of the power $p$ of the infalling radiation.  Indeed, the scalar energy density (\ref{eq:energyscalings}) just grows linearly with time $v$.  This happens because the energy flux through $\M$ is time-independent.  It turn out that the energy flux $\mathcal S$ through $\M$ is approximately time-independent so long as $|d_+ \Psi|^2|_{\lambda = 0}$ is slowly varying compared to $e^{-\kappa v}$.
In particular, with this assumption Eqs.~(\ref{eq:rdotRN}) and (\ref{eq:constraintlove}) yield $\zeta \sim |d_+ \Psi|^2|_{\lambda = 0}$
and Eq.~(\ref{eq:scalarflux}) implies $\mathcal S \sim |d_+ \Psi|^2|_{\lambda = 0}/\zeta \sim 1$.
This suggests our analysis can be generalized to de Sitter spacetime, where 
$|d_+ \Psi|^2|_{\lambda = 0}$ decays exponentially instead of with a power law.
It would be interesting to study the scenarios found in \cite{Costa:2017tjc,Cardoso:2017soq,Luna:2018jfk}.  We leave this for future work.

We conclude by discussing the generalization of our analysis to two-sided black holes.  Two-sided black holes have both singular ingoing and outgoing branches of the CH, where the geometry effectively ends.  For weakly perturbed two-sided black holes, the \textit{outgoing} branch of the CH lies at $r_{\rm CH}(v) = r_- - \varepsilon$ where $\varepsilon \to 0$ characterizes the strength of the perturbations \cite{Dafermos:2012np}. 
Therefore, when integrating the equations of motion (\ref{eq:einsteinscalar}) inwards along some $v = {\rm const.}$ geodesic, one cannot integrate beyond $r = r_{\rm CH}$, since a singularity is encountered there.  Correspondingly, our analysis of spacelike singularities, which takes place at $0 \leq r < r_{\rm c} \ll r_-$, cannot be applied to weakly perturbed two-sided black holes. 
This is consistent with the results of Ref.~\cite{Dafermos:2012np}, 
where it was demonstrated that weakly perturbed two-sided black holes 
only contain null singularities on the CH.  

What happens when perturbations of two-sided black holes are large? 
With large perturbations there is no reason to expect $r_{\rm CH}(v) \approx r_-$.  Indeed, numerical simulations of two-sided black holes with large perturbations indicate the CH contracts to $r = 0$, at which point it meets a spacelike singularity \cite{Brady:1995ni,Burko:1997zy}.  If at late times $r_{\rm CH}(v) \ll r_{\rm c}(v)$, then our analysis in this paper should apply.  Namely, Price Law tails result in a growing cloud of scalar radiation forming at $r \lesssim r_{\rm c}$, which itself necessitates the existence of a spacelike singularity at late enough times with the curvature near $r = 0$ growing like (\ref{eq:Kdivorigin}).  We shall report on numerical simulations of such a scenario in an upcoming paper.

\section{Acknowledgments}%
This work is supported by the Black Hole Initiative at Harvard University, 
which is funded by a grant from the John Templeton Foundation. EC is also supported by grant CU 338/1-1 from the Deutsche Forschungsgemeinschaft.  We thank Amos Ori and Jordan Keller for many helpful conversations during the preparation of this paper. 


\appendix 

\section{Numerics}
\label{app}

For the purpose of bolstering our assumptions that i) the geometry at $r > r_-$ relaxes to the RN solution and ii) that $A' = \partial_\lambda A$ varies smoothly across $r_-$, here we present a numerical solution to the Einstein-Maxwell-Scalar system.
For numerical simulations we have found it convenient to change radial coordinates from the affine parameter $\lambda$ to areal radius $r$.
The metric then takes the Bondi-Sachs form \cite{Madler:2016xju}
\begin{equation}
ds^2 = e^{2 B} [-2 V dv^2 + 2 dv dr] + r^2 (d\theta^2 + \sin^2 \theta d\phi^2).
\end{equation}
We numerically solve the Einstein-Maxwell-Scalar equations of motion using the methods detailed in \cite{Chesler:2013lia}.  We employ pseudospectral methods with domain decomposition and adaptive mesh refinement in the radial direction. For initial data we set  
\begin{equation}
\Psi = e^{-r^4/w^4},
\end{equation}
with $w = 1/2$. The mass and charge of the geometry were chosen to be $M = 0.9$ and $Q = 0.78$.  With these parameters $r_- = 0.45$ and $\kappa = 2.2$.  We then evolve the system from time $v = 0$ to $v = 4.1$.

\begin{figure}[h]
	\includegraphics[trim= 200 0 0 50 ,clip,scale=0.15]{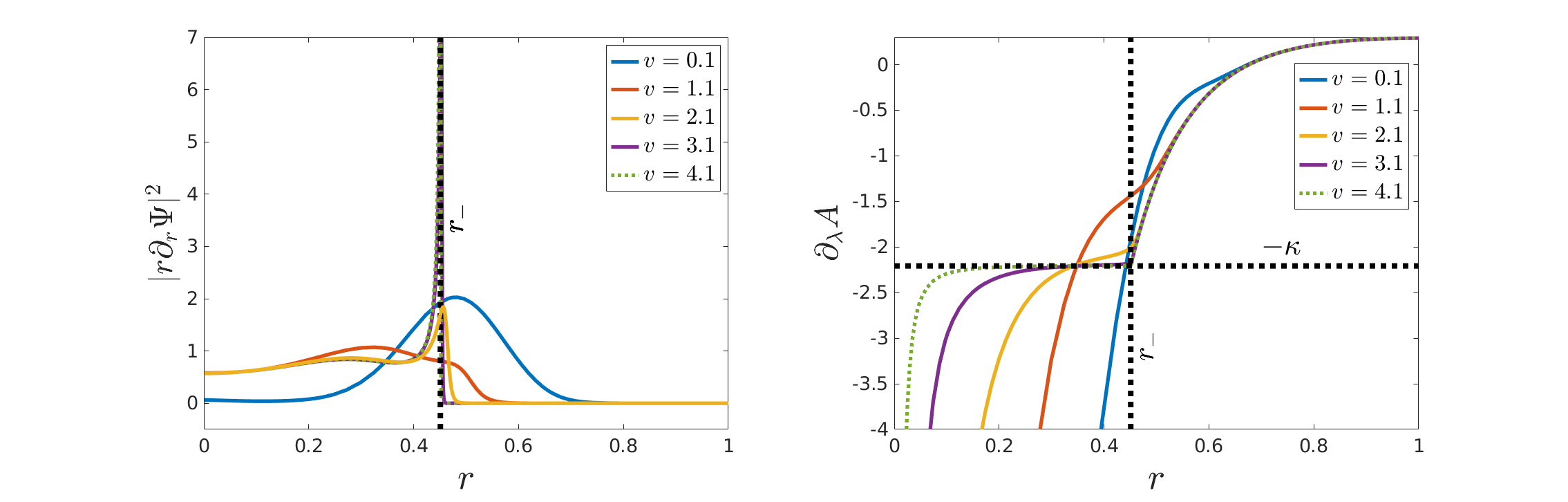}
	\caption{Left: the scalar energy density $r \mathcal E = |r \partial_r \Psi|^2$ at several times.  The scalar energy density becomes localized to a ball whose surface approaches $r_-$ at late  times.  By Birkoff's theorem, the geometry at $r > r_-$ must approach the RN solution.  Right: $\partial_\lambda A$ at several times.  At $r > r_-$ $\partial_\lambda A$ approaches its RN limit.  At $r < r_-$, $\partial_\lambda A$ approaches $-\kappa$ over an increasing large domain as $v$ increases.
	}
	\label{fig:appfig}
\end{figure}

In the left panel of Fig.~\ref{fig:appfig} we plot $r \mathcal E = |r \partial_r \Psi|^2$ at several times.  Initially the support of $|r \partial_r \Psi|^2$
extends beyond $r_-$.  However, as $v$ increases $|r \partial_r \Psi|^2$ becomes localized to a ball whose surface approaches $r_-$.  By Birkoff's theorem, the geometry at $r > r_-$ must therefore approach the RN solution.  Note derivatives of the scalar field at $r = r_-$ grow with time.

$\partial_\lambda A$ is related to 
$B$, $V$ and the electric field $E$ via
\begin{equation}
\frac{\partial A}{\partial \lambda} = 2 (\partial_v + V \partial_r) B + \frac{1}{2 r} e^{2B}(1 - 4 \pi r^2 E^2) - \frac{V}{r}.
\end{equation}
In the right panel of Fig.~\ref{fig:appfig} we plot $\partial_\lambda A$
at the same times shown in the left panel.  At $r > r_-$, $\partial_\lambda A$ approaches the associated RN expression as $v$ increases, with $\partial_\lambda A|_{r = r_-} \to - \kappa.$  Moreover, at $r < r_-$ we see that $\partial_\lambda A$ approaches $-\kappa$ over an increasing large domain as $v$ increases.
 
\section*{References}

\bibliography{refs}%

\bibliographystyle{utphys}
\end{document}